\documentclass[aps]{article}
\usepackage{geometry}              
\usepackage{graphicx}

\usepackage{amssymb}
\usepackage{epstopdf}
\usepackage{amsmath}

\usepackage{amsfonts}
\usepackage{bm}
\usepackage{hyperref}

\DeclareMathAlphabet{\mathpzc}{OT1}{pzc}{m}{it}

\title{The Transient Neutral Flux in Plasma: An Explanation of Heating for the Solar Corona?}  
\author{Clifford Chafin}
\author{Clifford Chafin\\\ \small{Human Innovations LLC,
506 Hamlet Ave.,	 Carolina Beach,	 North Carolina	 28428	 USA} \thanks{cechafin@ncsu.edu}}

\begin{document}
\maketitle
\begin{abstract}
In this short note, we discuss a mechanism for the transport of energy, momentum and dipole moment via transient neutral carriers in plasma.  This gives a way to rapidly convert bulk hydrodynamic flow energy into thermal energy over a distance of several mean free paths.  In the transition region of the solar corona we estimate various processes and their potential to introduce the high energies needed to to reach the $2\times10^{6}$K observed there.  It implies that kinetic methods may be essential for modeling the corona and that there are more gentle but still robust means than reconnection to relax magnetic fields in plasmas.  
\end{abstract}

The temperature of the solar corona is an enduring puzzle of plasma physics.  The photosphere is a mere 6000K.  This rises through the chromosophere to near 20,000K then undergoes a dramatic increase to 2x10$^{6}$K at the corona \cite{Priest}.  This seems completely at odds with thermodynamic models where monotonic cooling is expected as one gets further from the source of thermal energy.  
There are numerous attempts to explain the dramatically high temperatures in the solar corona \cite{Erdelyi}.  It is not our purpose to review the literature here or propose an elaborate model.  We only intend a note that discusses some of underlying assumptions on the local continuum mechanical treatments of plasmas.  In the course of this, we will give some suggestive justification for strong heating at the hydrodynamic-collisionless transition due to transient neutral bound pairs and propose origins of sources and sinks of magnetic flux in place of the usual models of reconnection.  

The Vlasov-Maxwell equations are modifications of the Boltzmann kinetic equations to describe plasmas and their interaction with electromagnetic fields.  Since they are kinetic equations they do not explicitly assume thermodynamic equilibrium or smooth data.  However, they do have some implicit assumptions that limit their applicability for the case of long range neutral fluxes.  They 
have the form 
\begin{align}
\frac{\partial f_{a}}{\partial t}+v_{a}\cdot\nabla f_{a}+q_{a}(E+(v\times B ))\cdot\nabla_{p}f_{a}&=0\\
\nabla\cdot E&=\rho\\
\nabla\times B&=0\\
\end{align}
\begin{align}
\rho&=\int d^{3}p (q_{i}f_{i}+q_{e}f_{e})\\
j&=\int d^{3}p (q_{i}f_{i}+q_{e}f_{e})(p/m)
\end{align}
where the index ``a'' corresponds to the electrons or ions in the plasma.  
The Boltzmann collision term is neglected based on the assumption that the major forces on the charges is due to bulk averaged electromagnetic ones.  In the frozen-in low density approximation the charges execute small circles and spirals about the magnetic field lines when $E=0$ and otherwise are drawn across them in an advancing spiral.  In a thermodynamic/hydrodynamic limit, there are particle interactions, mediated by the short range Coulomb and internal radiation fields, that drive thermalization of the local distribution and create hydrodynamic forces that are not described by this model.  
Neutral particles largely ignore the regionally averaged electromagnetic fields and the collision term becomes central to the dynamics.  The dominant effect of electromagnetism is in the high frequency radiation that is sufficient to break the bound pairs and convert them back to ions in the plasma.  

Realistic plasmas have both charges and neutral bound states that allow equilibration and interaction between these.  Near the surface of the sun, a nontrivial fraction of the charges are in such bound pairs.  Recombination is the result of proximity and the cross section of particle pairs.  On the other hand, pair breaking can be driven by collisions with other neutral particles, near collisions with ions (where the local rather than regional average fields matter) and photon flux.  The photon flux is difficult to model.  The photons that matter are those with energies and cross sections large enough to cause ionizing excitations and near the surface of the sun, where strong density and temperature gradients exist, the photon flux may be predominatly from regions where the gas/plasma temperature is significantly different.  This is but one source of strong nonlocality in the dynamics so that the continuum approximation implicit in hydrodynamic models should not be expected to hold.  Worse yet, the important transport of neutral pairs over distances much longer than the scale of variations of the magnetic and plasma flows suggests that this situation will be essentially nonperturbative.  

The Vlasov-Maxwell equations work best for cases where the magnetic field lines are ``frozen in'' the plasma.  Thus, if we have a parcel of plasma moving with sufficient momentum in a magnetic field, the parcel will polarize initially by counterrotating motions of the charges so that the edges pick up opposite charge and the $v\times B$ forces are cancelled by the induced electric ones.  Thermal equilibration will damp oscillations and lead to a persisting field.  This does not depend on the mean-free path or particle separation being smaller than the gyroradius.  Comparing with the motion of an isolated moving charge, the magnetic field simply freezes the motion into a circle so that only translation can occur along the B-lines in spirals.  One could add in more charges at low enough density so that this still occurs.  Eventually, at high enough density, the long range Coulomb forces will build up and polarization will allow the net translation of the parcel perpendicular to the external magnetic field.  

For the inclusion of a transient low density (i.e.\ nonthermalizing) neutral flux into the Vlasov equations, one must consider several effects.  It must include photons and so have a way to dynamically generate black or gray body distributions from the plasma distribution itself.  This is not a simple task.  It must include the change in magnetic source effects from the vanishing of counter rotating pairs of charges that, in turn, apply forces and torques on the charges generating the fields felt by this pair.  When these break up, we must have a source of magnetic flux and the forces and torques they generate.  If these new charges have long life compared to the thermalizing interaction times with other particles or the ambient radiation field then we can assume they thermalize without including the details of how this process occurs.  

Let us now make this more precise.  For simplicity we only consider a neutral plasma of Hydrogen gas.  Consider a relatively higher density region of plasma at temperature $T_{1}$ where with magnetic field $B_{1}$, density $\rho_{1}$ and flow velocity $v_{1}$ (parallel to $B_{1}$).  The fraction of, possibly excited, H-atoms are dragged along with this plasma due to collisions.  The charged particles move in spirals about the magnetic field in opposite motions depending on their charge.  Since mass and the gyroradius of the protons is much larger than that of the electrons we can approximate the local linear and angular momentum, kinetic energy and magnetic moment of a charged pair to be due to the protons alone.  

Assume we are in a region of large density gradient so that a fraction of the neutral pairs can travel long distances through the plasma into the lower density region with magnetic field $B_{2}$ and flow velocity $v_{2}$.  The bound pairs then, by collision with ambient photons, other neutral pairs or near interactions with ions, are forced to dissociate into ions that then circulate about the new local magnetic field.   To be specific about the local changes mediated by these pairs we need to know something about how they are created.  If the local temperature is low compared to the kinetic energy then a there is no restriction on the relative velocities of the electrons and protons that bind.  If the temperature is much larger then only pairs that have similar velocities can efficiently pair up.   Near the optical surface of the sun, $kT\approx5$eV $<$ Ry$\approx$13.6eV so we assume that the former case holds as a rough approximation.  

The typical thermal energy of electron and protons is both $E=\frac{3}{2}kT_{1}$.  The net angular momentum this represents is 
\begin{align}
L\approx m_{p}v_{th}r_{g}=\frac{m_{p}^{2}v_{th}^{2}}{qB_{1}}
\end{align}
On average, pairs pick up a net kinetic motion form the flow they are in so the momentum is $P\approx m_{p}v_{flow}$.  The magnetic moment is 
\begin{align}
\mu\approx qv_{th}\pi r_{g}^{2}=\pi\frac{m_{p}^{2}v_{th}^{3}}{q B_{1}^{2}}
\end{align}
Here $v_{th}\approx\sqrt{3kT_{1}/m_{p}}$.  The rate that such pairs vanish from this region gives the local forces and torques on the plasma generating the fields and changes in their local contribution to them.  

The pairs travel to the new region and dissociate.  These now exhibit orbits about the new magnetic field axis $\hat{B}_{2}$ and transfer momentum and torques to the field producing plasma here.  This will be a function of $|B_{1}\times B_{2}|$.  If these flows are perpendicular, as in the case of a spicule or other upwardly rising plume of charges and a region with fields parallel to the sun we can assume the energy loss is maximal.  Each pair then transports an energy of $KE\approx \frac{1}{2}m_{p}v_{flow}^{2}$.  The magnetic field has also undergone a discontinuous transfer in its sources undergoing a change at the origination point of 
\begin{align}
\Delta\vec{\mu}=-\pi\frac{m_{p}^{2}v_{th}^{3}}{q B_{1}^{2}}\hat{B}_{1}
\end{align}
and a change at the destination of 
\begin{align}
\Delta\vec{\mu}=\pi\frac{m_{p}^{2}v_{th}^{'3}}{q B_{2}^{2}}\hat{B}_{2}
\end{align} 
This gives a mechanism beyond reconnection to release magnetic energy and do so in a more gradual fashion.  Although we will not pursue it here, it may also provide a mechanism for proto-stars to transfer angular momentum outwards especially in the cooler regions where neutral carriers are more numerous.

For a concrete example, consider the Hydrogen gas atoms near the top of a spicule or moving at the equatorial surface velocity of the sun that is then transferred out of its frozen in plasma to a region that has a magnetic field approximately stationary with respect to the center of mass of the sun.  The surface velocity of the sun at the equator is $\sim2$km/s.  A spicule has vertical velocity $\sim20$km/s and can reach 10's of thousands of km up into the corona, where the mean free path is many kilometers in length.  Microspicules can have velocities of 150km/s \cite{Erdelyi}.  Let us consider a thermalized neutral pair that originates in such a location and dissociates in one of the magnetically stationary regions relative to the sun.  The direct conversion of translational kinetic energy of the flow gives the temperatures as a function of various collectively moving flows and other sources as shown in table \ref{tab}.

\begin{table}
\begin{center}
    \begin{tabular}{| l | l |}
    \hline
    Source Region of TNF & Final Temp (K) \\ \hline\hline
    Solar Surface & 160  \\ \hline
    Typical Spicule & 16,000 \\ \hline
    Microspicule & $9\times10^{5}$ \\ \hline
    CME & $10^{7}$ \\ \hline
    Coronal X-ray Jet & $1.6\times10^{6}$ \\ \hline
    \end{tabular}
    \end{center}
    \caption{Typical values for equilibration temperature from TNFs into magnetic regions stationary relative to the sun.}\label{tab}
\end{table}

This process has no way to generate higher temperatures than the kinetic energy of the neutral fluxes at their source so it is not just a matter of determining energy flux in determining temperature of the heating.  The originating flow velocity is the limiting factor.  Strong coronal heating has been detected at regions of strong spicule activity.  This generally seems too weak an effect to generate such temperatures unless there are much faster leading edge components to these eruptions.  If TNF is the source of coronal heating it seems most likely to be from coronal X-ray jets or extreme spicule events.  Given the temperature predictions and frequency and large scale occupied by X-ray jets they seem like the more promising candidate.  However, as we have observed through sprites in lightning storms on earth, there can be high energy but low power events on top of very energetic ones.  A more rigorous approach to modeling this effect may lead to more definite predictions but continuum methods look hopeless so a kinetic approach seems unavoidable.

\end{document}